\documentclass{Interspeech}

\usepackage{balance}
\usepackage{booktabs}
\usepackage{adjustbox}
\usepackage{array}
\usepackage{bm}
\usepackage{multirow}
\usepackage{algorithmic}
\usepackage{algorithm}
\usepackage{cite}
\usepackage{graphicx}
\usepackage{subcaption}
\usepackage{amsmath,amssymb,amsfonts}
\usepackage{url,times,booktabs, tabularx}
\usepackage[activate]{microtype}
\usepackage{comment}

\usepackage[activate]{microtype}
\newcommand{\sstitle}[1]{\smallskip\noindent\textbf{#1.\/}}

\newcolumntype{P}[1]{>{\centering\arraybackslash}p{#1}}
\newcolumntype{R}[1]{>{\raggedleft\arraybackslash}p{#1}}

\newcommand{\eg}{e.\,g.,\ }
\newcommand{\ie}{i.\,e.,\ }

\interspeechcameraready

\title{Breaking Resource Barriers in Speech Emotion Recognition via Data Distillation}

\author[affiliation={1}]{Yi}{Chang}
\author[affiliation={2}]{Zhao}{Ren}
\author[affiliation={3}]{Zhonghao}{Zhao}
\author[affiliation={4}]{Thanh Tam}{Nguyen}
\author[affiliation={3}]{Kun}{Qian}
\author[affiliation={2}]{Tanja}{Schultz}
\author[affiliation={1,5}]{Björn W.}{Schuller}

\affiliation{GLAM -- the Group on Language, Audio, \& Music}{Imperial College London}{United Kingdom}
\affiliation{Cognitive Systems Lab}{University of Bremen}{Germany}
\affiliation{School of Medical Technology}{Beijing Institute of Technology}{China}
\affiliation{}{Griffith University}{Australia}
\affiliation{Chair of Health Informatics, Klinikum rechts der Isar (MRI)}{Technical University of Munich (TUM)}{Germany}


\usepackage{comment}

\begin{document}

\maketitle

\begin{abstract}
    
    Speech emotion recognition (SER) plays a crucial role in human-computer interaction. The emergence of edge devices in the Internet of Things (IoT) presents challenges in constructing intricate deep learning models due to constraints in memory and computational resources. Moreover, emotional speech data often contains private information, raising concerns about privacy leakage during the deployment of SER models. To address these challenges, we propose a data distillation framework to facilitate efficient development of SER models in IoT applications using a synthesised, smaller, and distilled dataset. Our experiments demonstrate that the distilled dataset can be effectively utilised to train SER models with fixed initialisation, achieving performances comparable to those developed using the original full emotional speech dataset.
\end{abstract}

\section{Introduction} 
Speech, a distinctive human capability, facilitates the conveyance of a wide range of emotional nuances~\cite{wani2021comprehensive} (\eg happiness, sadness, anger, fear, etc.). Identifying emotions from speech is a meaningful and challenging task~\cite{schuller2018speech, chang2024staanet}, as internal emotional states are often hidden in speech and difficult for humans to recognise~\cite{al2015emotion}. Over the past decades, speech emotion recognition (SER) aims to automatically identify emotional states from speech signals through signal processing and artificial intelligence approaches. An escalating number of human-computer interaction applications, such as intelligent call centres~\cite{app122110951} and computer games~\cite{8925464}, are evidencing a burgeoning need for SER.

With the advancement of computing resources, numerous deep learning (DL) methodologies have been successfully applied to the task of SER~\cite{madanian2023speech, ZHAO2019312, 8817913, 10141880}, including convolutional neural networks (CNNs) with the input of time-frequency representations~\cite{ZHAO2019312, 8817913} and end-to-end models (\eg wav2vec 2.0~\cite{baevski2020wav2vec}, HuBERT~\cite{hsu2021hubert}, and WavLM~\cite{chen2022wavlm}) fed with raw audio samples. Currently, the integration of DL techniques for SER into the Internet of Things (IoT) has spurred the development of numerous innovative personalised applications, such as smart homes~\cite{9352018}, health states monitoring~\cite{9133298}, and intelligent transportation systems~\cite{khalil2021deep}. In this context, applying DL to IoT frameworks shows two major challenges. 
First, the constraints of DL models imposed by memory consumption continue to impede their advancement on edge devices, which often have limited onboard resources~\cite{chen2020deep}.
Second, considering that speech data encompasses a wide array of personal information~\cite{hashem2023speech}, recent research findings have demonstrated the potential vulnerability of DL algorithms to privacy breaches~\cite{al2020survey,yin2017mobi}.

To address the aforementioned two challenges, data distillation emerges as a promising solution by extracting representative information to form a refined, smaller dataset from a larger one. 
Specifically, data distillation generates a compact dataset synthesised from the entirety of the original data, enabling that models trained on the smaller dataset perform on par with those on the complete one~\cite{wang2018dataset}. 
Additionally, leveraging synthesised smaller-scale datasets for subsequent model training further diminishes the likelihood of privacy leakage~\cite{sachdeva2023data}. 
Data distillation has emerged as a vibrant research area in the field of machine learning with diverse applications (\eg image classification~\cite{cazenavette2022dataset}). However, the application of data distillation remains limited in SER. To the best of our knowledge, this is the first attempt to construct a data distillation framework specifically for SER.

In this paper, we present a data distillation framework to synthesise a compact dataset for SER by constructing a similar model parameters with fewer iterations~\cite{cazenavette2022dataset}. The main contributions of this work can be summarised as follows: First, the proposed approach can effectively reduce the dataset size to approximately 15\,\% of the original one, conserving memory space while maintaining performance. Second, the required training iterations on the distilled dataset are less than those on the original data. 
Third, our framework has potential in protecting user identification against malicious exploitation in the SER domain. The code will be available online upon the paper's acceptance.

 \begin{figure*}[]
    \centering
    \includegraphics[trim={0 10 0 0},clip,width=.65\textwidth]{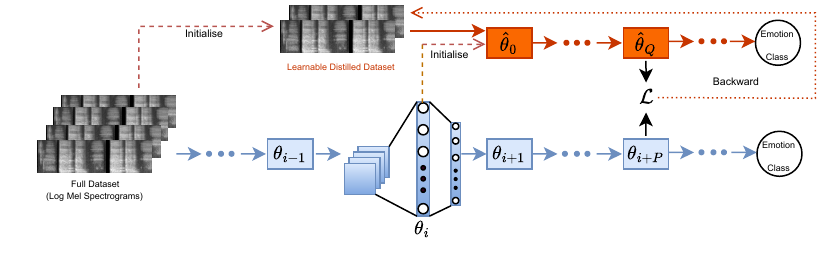}
    \vspace{-5pt}
    \caption{Overview of the proposed data distillation framework for SER. The blue arrows indicate the teacher networks' training on the original, complete dataset, whereas the red arrows display the student networks' training on the learnable distilled dataset.}
    \label{fig:workflow}
\end{figure*}

\section{Related Work}
Our work addresses memory constraints on edge devices by reducing model parameters.
For instance, model pruning can effectively reduce model size and computational complexity by strategically removing some model parameters~\cite{han2015learning}. Parameter quantisation reduces the storage space of a model by decreasing the number of bits needed for each parameter~\cite{gholami2022survey}. Model distillation~\cite{hinton2015distilling} is a method that transfers knowledge from a complex teacher model to a lightweight student model, making it well-suited for edge computing and IoT applications~\cite{hinton2015distilling}. Compared to the aforementioned methods for model size reduction, data distillation entails transferring knowledge from a large-scale dataset to a smaller one, enabling models trained on the compact dataset to match the performance of those trained on the original dataset~\cite{wang2018dataset}. Data distillation has been widely researched within ML (\eg continual learning~\cite{li2023memory}, federated learning~\cite{sucholutsky2021soft}, and neural architecture search~\cite{nguyen2021dataset}). Most existing research focuses on image datasets, like MNIST~\cite{yann1998mnist} and CIFAR~\cite{krizhevsky2009learning}. 
Studies~\cite{mohapatra2022speech} and~\cite{you2020towards} have sought to apply dataset distillation techniques in the domains of speech recognition and question-answering systems, respectively. Inspired by the these studies, data distillation~\cite{cazenavette2022dataset} is investigated herein to reduce data size for SER in our work.

To protect users' privacy in the context of IoT, federated learning aims to keep users' data on local clients (\ie edge devices) while only uploading model parameters to the cloud server. This approach helps to safeguard privacy by minimising exposure during client-server communication as well as in the cloud, though it requires the exchange of complex model parameters~\cite{sachdeva2023data}. Data distillation can protect users' privacy by generating dataset summaries from few number of data samples~\cite{sachdeva2023data}. 
Moreover, data distillation has been shown to be effective in differential privacy~\cite{dong2022privacy}. Furthermore, integrating federated learning with data distillation, which involves uploading synthesized data rather than model parameters, not only reduces the communication load between clients and the server but has also been shown to outperform traditional federated learning with model synchronisation~\cite{hu2022fedsynth,liu2022meta}.

\section{Methodology}

Based on the original, complete and full SER dataset $D_{\text{org}}$, dataset distillation in this work aims at generating a smaller but effective, synthetic training dataset $D_{\text{syn}}$. As depicted as in Figure~\ref{fig:workflow}, log Mel spectrograms extracted from the audio samples in $D_{\text{org}}$ are fed into a convolutional neural network (CNN) for emotion classification, where the snapshot of model parameters after each iteration are stored into a \emph{teacher trajectory}. When conducting the dataset distillation, we first initialise the trainable $D_{\text{syn}}$ by randomly selecting a certain number of log Mel spectrograms per class (IPC) from $D_{\text{org}}$. Second, a \emph{student trajectory} is initialised with a randomly chosen iteration (\ie $i$ in Figure~\ref{fig:workflow}) from a \emph{teacher trajectory}. Afterwards, the \emph{student trajectory} is trained on the $D_{\text{syn}}$ for several iterations (\ie $Q$ in Figure~\ref{fig:workflow}). During the dataset distillation, the values of the log Mel spectrograms are updated according to the distance between $\theta_{i+P}$ and $\hat{\theta}_{Q}$, where $P$ is a pre-defined number of epochs along the \emph{teacher trajectory}. In this way, the knowledge of all emotional information in the full dataset is transferred from the \emph{teacher trajectory} to the generated $D_{\text{syn}}$. In Section~\ref{sec:teacher_cnn} and \ref{sec:student_cnn}, we introduce the SER models applied for detailed trajectories construction. In Section~\ref{sec:distil}, the loss function in the training of $D_{\text{syn}}$ is described.

\subsection{Teacher Trajectories}
\label{sec:teacher_cnn}
In this work, the trajectory means a sequence of a model's parameters. When training the CNN models on $D_{\text{org}}$, we store the snapshot parameters of CNN models after each epoch as $\theta_{i}$, where $i$ represents the current iteration number. A \emph{teacher trajectory} is defined as $\Theta = \{ \theta_1, \theta_2, \ldots, \theta_e \}$, where $e$ denotes the number total iterations. In this work, we generate a set of \emph{teacher trajectories} $\mathcal{T} = \{\Theta_1, \Theta_2, \ldots, \Theta_t\}$ for robustness.

In considering deployment on edge devices, log Mel spectrograms are preferred over raw audio samples due to their lower storage resource requirements~\cite{10096757}. Furthermore, CNNs have demonstrated their proficiency in extracting effective representations from log Mel spectrograms~\cite{ren2020generating, chang2021covnet}.
Therefore, we evaluate the effectiveness of our proposed framework for SER using three classic CNN models: CNN-6, ResNet-9, and VGG-15. These CNN models are also suitable for deployment on devices with limited computing resources~\cite{10096757}. The detailed model architectures are described in Section~\ref{sec:exp_set}.

\subsection{Student Trajectories}
\label{sec:student_cnn}
Similarly, a \emph{student trajectory} denotes also a sequence of model's parameters trained on the $D_{\text{syn}}$.
For faster distillation process and robustness, at each distillation step, we first randomly choose one \emph{teacher trajectory} $\Theta_j$ from $\mathcal{T}$ and then randomly chose the $i-th$ snapshot of a model's parameters $\theta_i$ from $\Theta_j$. However, due to the diminishing informational value (\ie model parameter changes) in later iterations during the training, we set an upper bound $e^{+}$ for the value $i$ in the $\theta_i$ for the \emph{student trajectory} initialisation.

At each distillation step, following the student trajectory's initialisation, there are in total of $Q$ numbers of updates based on the classification loss of the synthetic data. Specifically, considering the 
memory constraints, the synthetic dataset is partitioned into batches, and gradient descent updates are performed at each time according to the cross-entropy loss $\mathcal{L}'$ calculated from the current batch of synthetic data, indicated as follows: 
\vspace{-5pt}
\begin{equation}
\begin{aligned}
\hat{\theta}_{n+1} = \hat{\theta}_{n} - \alpha \nabla \mathcal{L}'((b_{n}); \hat{\theta}_{n}),
\end{aligned}
\label{eq:grads}
\end{equation}
where $n \in \{0, 1, \ldots, Q-1\}$ indicates the current iteration number and $b_{n}$ is the sampled batch of synthetic data, and $\alpha$ is a trainable learning rate
to automatically adjust the magnitude of iterations $P$ and $Q$ for the student trajectories at each distillation step. After a certain number of distillation steps, the final distilled data $D_{\text{syn}}$ is generated. 

\subsection{Training of Distilled Data}
\label{sec:distil}

After $Q$ updates to the student trajectory, with the snapshot of parameters $\theta_{i+P}$ retrieved from $\Theta$ and the parameters $\hat{\theta}_{Q}$, the distilled log Mel spectrograms are updated according to the normalised squared $l$-norm distance as follows:
\vspace{-5pt}
\begin{equation}
\begin{aligned}
\mathcal{L} = \frac{\lVert \hat{\theta}_{Q} - \theta_{i+P} \rVert_l^2}{\lVert \theta_{i} - \theta_{i+P} \rVert_l^2}, 
\end{aligned}
\label{eq:loss}
\end{equation}
where $Q \ll P$, showing the efficiency of the distillation process. This loss function directly encourages 
a similar student trajectory $\hat{\theta}_{Q}$ to the teacher trajectory $\theta_{i+P}$ trained on the $D_{\text{org}}$ in the parameter space with much less iteration steps. The normalisation in the above loss function ensures the continuity and effectiveness of the learning signal across different training stages by maintaining the relative scale of updates and preventing the potentially diminishing gradient issue, especially in the later training epochs~\cite{cazenavette2022dataset}.

\section{Experiments and Results}

\subsection{Dataset}

\sstitle{DEMoS} 
The Database of Elicited Mood in Speech (DEMoS)~\cite{parada2019demos} is an Italian speech corpus consisting of 7.7\,hours of audio recordings from $68$ participants ($45$ males and $23$ females). This dataset not only includes 332 neutral speech samples but also features a collection of $9,365$ audio samples categorised into seven emotional states: anger, disgust, fear, guilt, happiness, sadness, and surprise. Same to prior studies~\cite{ren2020enhancing, 10096757, 10094895} on the DEMoS dataset, this work excludes the minority neutral class, focusing on the $9,365$ emotional speech recordings. These $9\,365$ samples are then allocated into training ($40$\,\%), validation ($30$\,\%), and test ($30$\,\%) sets, adhering to a speaker-independent approach. The detailed data distribution of the DEMoS dataset is outlined in Table~\ref{tab:demos_distribution}. 

\begin{table}[ht]
\centering
\caption{Emotion distribution of the DEMoS dataset.}
\vspace{-10pt}
\begin{adjustbox}{width=0.75\columnwidth}
\begin{tabular}{l|R{.7cm}R{.7cm}R{.7cm}|R{.7cm}|r}
\toprule
\# & Train & Val & Test & $\bm{\sum}$ & F / M \\ \hline
\multicolumn{1}{l|}{Speaker} & 27 & 25 & 16 & 68 & 23 /45 \\
\hline
\multicolumn{1}{l|}{Anger} & 586 & 531 & 360 & 1,477 & 400 / 729 \\
\multicolumn{1}{l|}{Disgust} & 666 & 608 & 404 & 1,678 & 596 / 1,082 \\
\multicolumn{1}{l|}{Fear} & 461 & 404 & 291 & 1,156 & 524 / 871 \\
\multicolumn{1}{l|}{Guilt} & 453 & 395 & 281 & 1,129 & 415 / 741 \\
\multicolumn{1}{l|}{Happiness} & 561 & 471 & 363 & 1,395 & 516 / 961 \\
\multicolumn{1}{l|}{Sadness} & 606 & 543 & 381 & 1,530 & 349 / 651 \\
\multicolumn{1}{l|}{Surprise} & 396 & 358 & 246 & 1,000 & 532 / 998 \\ \hline
$\bm{\sum}$ & 3,729 & 3,310 & 2,326 & 9,365 & 3,332 / 6,033 \\ 
\bottomrule
\end{tabular}
\label{tab:demos_distribution}
\end{adjustbox}
\vspace{-10pt}
\end{table}

\subsection{Experimental Settings}
\label{sec:exp_set}
\sstitle{Evaluation Metrics}
The Unweighted Average Recall (UAR), alongside accuracy (\ie weighted average recall), serves as the primary evaluation metric in alignment with previous studies on the DEMoS~\cite{10096757, 10094895} corpus, aiming to mitigate the issue of class imbalance.

\sstitle{Models Architecture}
The CNN-6 model consists of four convolutional layers with output channels of $64$, $128$, $256$, and $512$, each with a kernel size of $3 \times 3$. Each convolutional layer is followed by a local max pooling layer with a kernel size of $2 \times 2$. The VGG-15 model features five convolutional blocks: the first two blocks each contain two convolutional layers, while the latter three blocks each comprise three convolutional layers. The number of output channels in these blocks is set as $64$, $128$, $256$, $512$, and $512$, with all convolutional layers utilising $3 \times 3$ kernels. Following each convolutional block, a max pooling layer with a $2 \times 2$ kernel is employed. The ResNet-9 model initiates with a convolutional layer of $64$ output channels with a $7 \times 7$ kernel and a stride of $2$, followed by a max pooling layer with a $3 \times 3$ kernel. Subsequent to this are three ResNet blocks  with output channel numbers $128$, $256$, and $512$. These blocks leverage `shortcut connections' that combine identity mappings with the outputs from two consecutive $3 \times 3$ convolutional layers~\cite{ren2020generating}. For classification purposes, a global max pooling layer is utilised, followed by two fully connected (FC) layers to discern the contribution of each time-frequency bin. In this way, CNN-6, ResNet-9, and VGG-15 contain $4.44$\,M, $4.96$\,M, and $14.86$\,M parameters, respectively.

\begin{figure*}[!h]
    \centering
    \begin{subfigure}[b]{0.3\linewidth}
    \centering
        \includegraphics[width=1\linewidth]{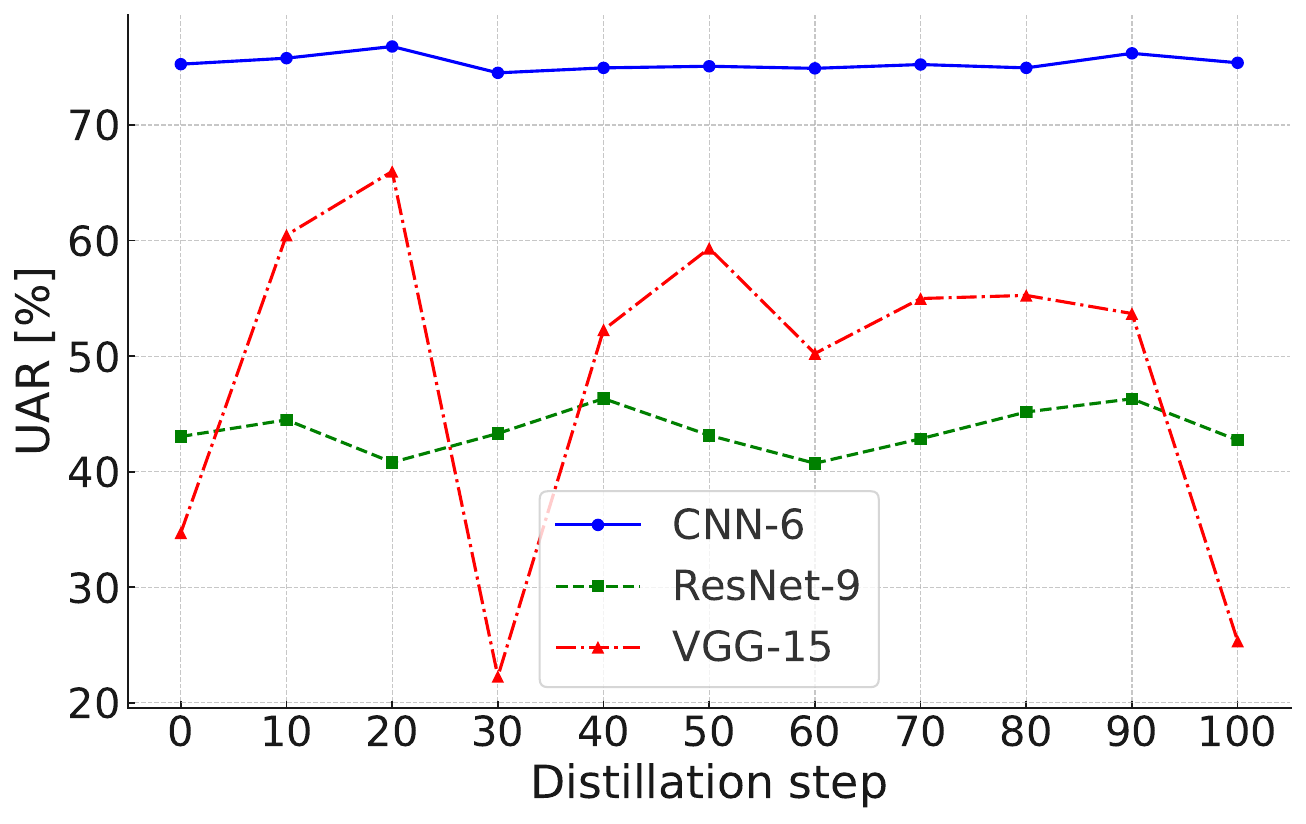}
        \caption{Trained on data distilled with CNN-6.}
        \label{fig: 1a}
    \end{subfigure}
    \hspace{0.1em}
    \begin{subfigure}[b]{0.3\linewidth}
    \centering
        \includegraphics[width=1\linewidth]{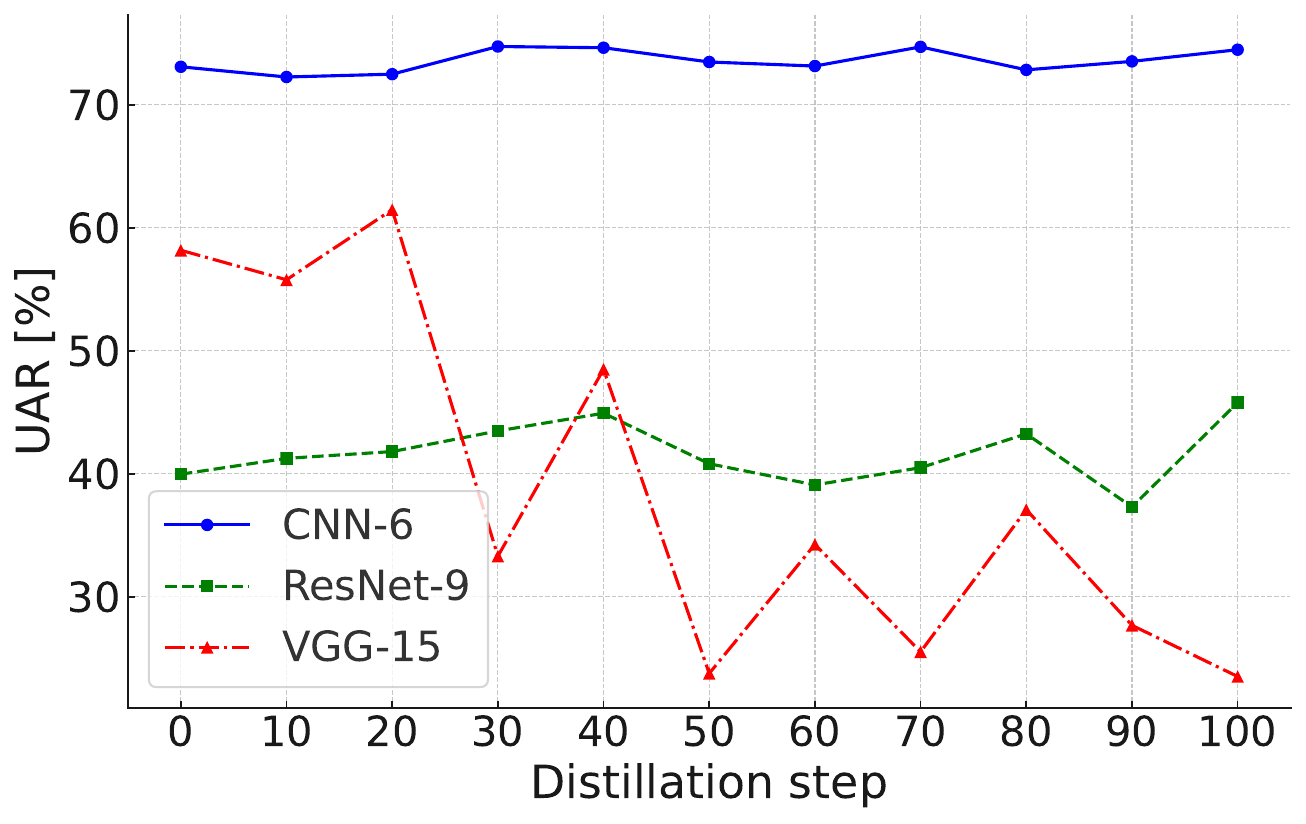}
        \caption{Trained on data distilled with ResNet-9.}
        \label{fig: 1b}
    \end{subfigure}
    \hspace{0.1em}
    \begin{subfigure}[b]{0.3\linewidth}
    \centering
        \includegraphics[width=1\linewidth]{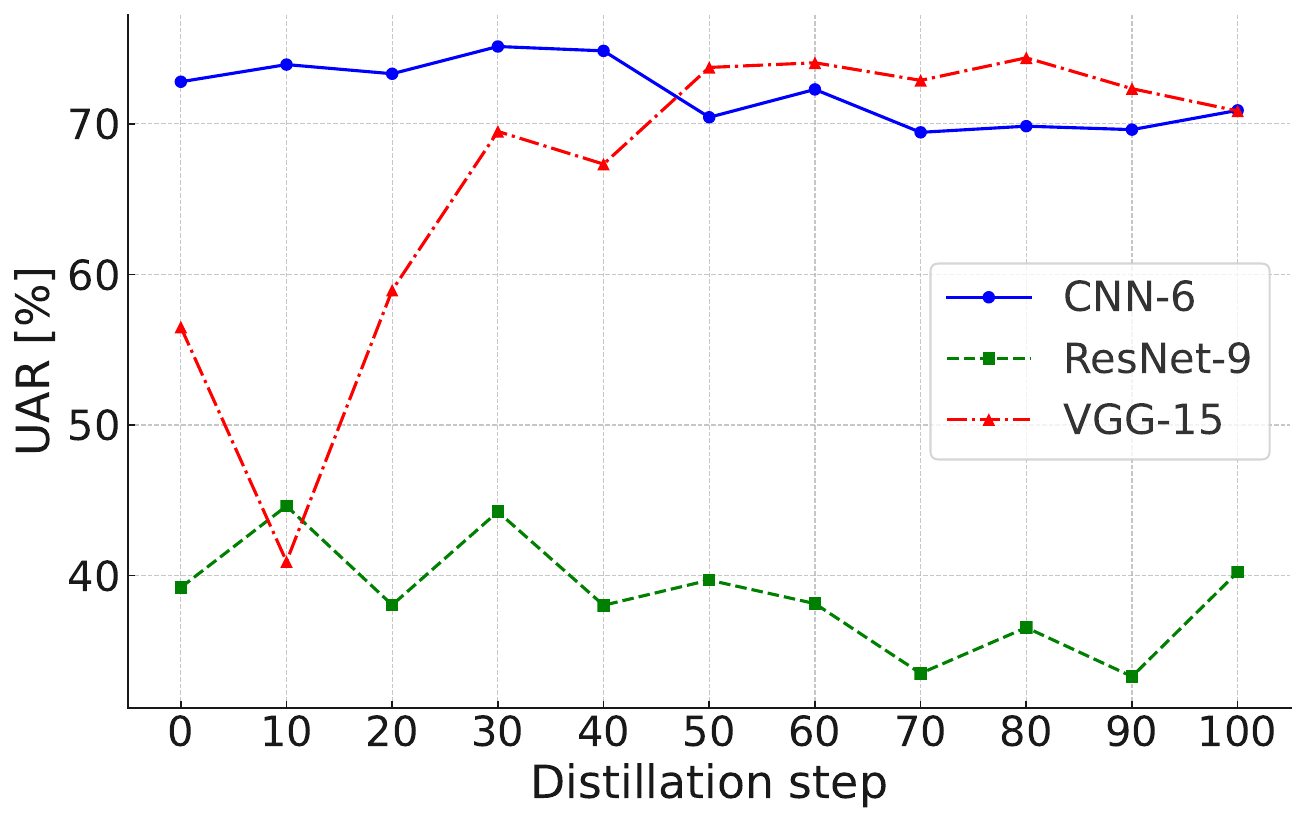}
        \caption{Trained on data distilled with VGG-15.}
        \label{fig: 1c}
    \end{subfigure}
    \vspace{-10pt}
    \caption{Comparison of the performance (UAR [\%]) of the models on the DEMoS validation datasets, when IPC is set as $150$.}
    \vspace{-15pt}
    \label{fig:dis_step}
\end{figure*}

\sstitle{Implementation Details} 
In this work, all audio samples in the DEMoS dataset are re-sampled into $16$\,kHz for faster processing. Moreover, we unify the duration of all audio samples to match that of the longest clip by trimming excessive portions and self-repeating segments of shorter ones. We extract log Mel spectrograms from DEMoS audio samples as features for models development. Specifically, the sliding window, overlap, and Mel bins are set as $512$, $256$, and $64$ time frames, respectively. As a result, the obtained log Mel spectrograms are shaped as ($373$, $64$), with the $373$ representing the time axis and $64$ describing the number of Mel frequency bins. The developing of the models employs an Adam optimiser with an initial learning rate of $1e-3$ and stops after $50$ epochs. The learning rate decreases by $30$\,\% after every $5$ epochs. 

In our distillation setup, we define the number of teacher trajectories as $N_{\Theta} = 5$ to ensure robustness against random initialisation of the student trajectories. We empirically set the maximum starting iteration $e^{+}$ to $30$ epochs when randomly selecting a parameter snapshot from $\Theta$. For the initialisation of trainable distilled data, we sample $\{1, 5, 10, 50, 100, 150\}$ log Mel spectrograms per class (IPC). The iteration number $Q$ applied to the student trajectory at each distillation step is fixed at $20$, while $P$ is set to $3$ epochs of iterations. The synthetic data is updated using an Stochastic Gradient Descent (SGD) optimiser with an initial learning rate set to $1000$. Concurrently, the learning rate $\alpha$, used in the gradient updates (as defined in Equation~\ref{eq:grads}), starts at $1e-3$ and is also adjusted using an SGD optimiser, with an initial learning rate of $1e-3$. The $l$ is set as $2$ in Equation~\ref{eq:loss}. The entire distillation process is iterated for $60$ cycles empirically. During the distillation process, the synthetic data is utilised every 10 distillation steps to develop three CNN models introduced in Section~\ref{sec:teacher_cnn} from scratch, which are subsequently evaluated on the validation and test datasets of DEMoS. Given the reduced size of the synthetic dataset, the total number of training epochs is reduced to $30$, while all other settings remain consistent with the development of the teacher trajectories.

\subsection{Results and Analysis}
\sstitle{Models developed on full dataset}
To ensure the efficacy of teachers networks in data distillation, we compare our performance with those of all state-of-the-art (SOTA) approaches on the DEMoS dataset, as shown in Table~\ref{tab:full_results}. CNN-6 achieves best validation UAR and best test UAR. Moreover, our approach outperforms most prior works on DEMoS dataset. Please note that the data split in~\cite{10094895} is different from our work and the work in~\cite{10094895} applies a more advanced wav2vec 2.0 model.

\begin{table}[ht]
\centering
\caption{Comparison of the performances [UAR \,\%] on the DEMoS dataset between our models (lower lines) and the state-of-the-art (SOTA). We present the mean $\pm$ standard deviation of accuracy and UAR values. }
\vspace{-10pt}
\begin{adjustbox}{width=1\columnwidth}
\label{tab:full_results}
\begin{tabular}{lcccc}
\toprule
Models & \multicolumn{2}{c}{Validation Dataset} & \multicolumn{2}{c}{Test Dataset} \\
\cmidrule(r){2-3} \cmidrule(r){4-5}
 & Accuracy & UAR  & Accuracy & UAR \\
\midrule
\textit{Model self-distillation~\cite{10094895}} &  - & 91.8 & - & 91.4 \\
\textit{VGG-16 with Adv~\cite{ren2020generating}} &  - & 87.5 & - & 86.7 \\
\textit{VGG-15 with NSL~\cite{10096757}} &  74.9 & 70.0 & 85.9 & 78.9 \\
\addlinespace 
CNN-6 & 88.9 $\pm$ 0.7 & \textbf{88.8 $\pm$ 0.9} & 88.3 $\pm$ 0.6 & \textbf{88.3 $\pm$ 0.6} \\
ResNet-9 & 74.8 $\pm$ 0.9 & 74.6 $\pm$ 1.0 & 81.5 $\pm$ 0.4 & 81.1 $\pm$ 0.4 \\
VGG-15 & 86.9 $\pm$ 1.4 & 86.8 $\pm$ 1.3 & 87.1 $\pm$ 1.1 & 87.0 $\pm$ 1.1 \\
\bottomrule
\end{tabular}
\end{adjustbox}
\end{table}

\sstitle{Models developed on the distilled dataset}
Considering the memory constraints on IoT devices, we experiment with various sizes of synthetic datasets. 
While a higher IPC can enhance model performance, it also necessitates a larger dataset. Thus, we set the IPC values to a maximum of $150$ to ensure the distilled datasets remain sufficiently small.
Specifically, certain numbers of Log Mel spectrograms per class (IPC) are randomly chosen from the real original dataset and updated throughout the distillation process, with IPC values set as $\{1, 5, 10, 50, 100, 150\}$. These IPC values represents $0.1\,\%$, $0.5\,\%$, $1.0\,\%$, $5.0\,\%$, $9.9\,\%$, $14.9\,\%$ of the training plus validation data samples and $0.2\,\%$, $0.9\,\%$, $1.9\,\%$, $9.4\,\%$, $18.8\,\%$, $28.2\,\%$ of the training samples. 

For each IPC, we apply the DEMoS validation dataset to test the effectiveness of the distilled data after every $10$ distillation steps, up to $100$. Figure~\ref{fig:dis_step} presents the results when $IPC=150$. First, the performance of CNN-6 remains stable across the distillation steps, regardless of whether the synthetic data is guided by CNN-6, ResNet-9, or VGG-15, which may be attributed to its relatively simpler architecture and the adequacy of $150$ IPC for training CNN-6. Second, VGG-15 shows improved performance when trained on synthetic data distilled by VGG-15 itself, likely due to its more complex architecture. Third, as evidenced by Figure~\ref{fig:dis_step}, apart from the case with IPC $=100$, we observe a distillation step of $60$ appears optimal. Further distillation leads to no improvement or even a decrease in performance, possibly indicative of overfitting.

\begin{figure}[!t]
\centering
    \includegraphics[width=0.7\linewidth]{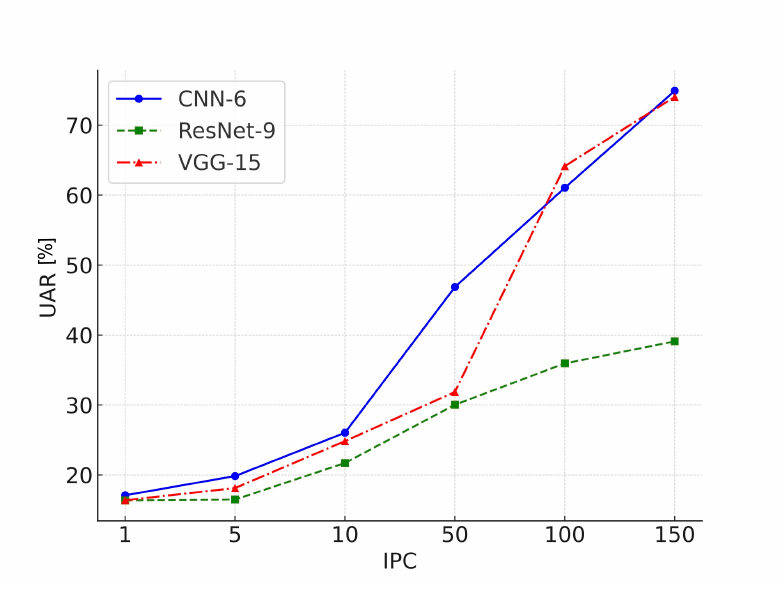}
    \vspace{-5pt}
    \caption{Analysis of the image per class (IPC) on the DEMoS validation dataset when the distillation step is fixed to $60$. The model type in data distillation and evaluation remains same.}
    \label{fig:ipc_vary}
    \vspace{-20pt}
\end{figure}

After fixing the total number of distillation steps to $60$, we examine the impact of IPC on the efficacy of the synthetic data used for training from scratch. In Figure~\ref{fig:ipc_vary}, there is a discernible trend showing that the performance of models trained on the synthetic distilled data improves with the increase of IPC. Notably, CNN-6 exhibits a more pronounced and rapid improvement, with considerable gains observed post an IPC threshold of $10$. Upon reaching an IPC of $150$, the performance of VGG-15 closely aligns with that of CNN-6 and their UARs exceeds $70\,\%$, outperforming one SOTA work~\cite{10096757} in Table~\ref{tab:full_results}. An IPC of $150$ means we train the model with $28.2\,\%$ of the original validation dataset. A comparable performance on the distilled data with IPC$=150$ has verified the effectiveness of the proposed approach, which is another reason we do not increase the IPC further.

In Table~\ref{tab:res}, we present the model performances trained on distilled data guided by their respective model architectures as well as others. With IPC set at 150, the train dataset constitutes $28.2\,\%$ and the combined train plus validation set accounts for $14.9\,\%$ of the original corresponding dataset. On the validation dataset, CNN-6 achieves a UAR of $88.8\,\%$ when trained on the full train dataset. Interestingly, even when trained on merely $28.2\,\%$ of the data samples, its UAR remains above $72.0\,\%$, peaking at 
$74.9\,\%$ when guided by its own architecture during the data distillation process. Similarly, ResNet-9 maintains a consistent validation UAR of around $40.0\,\%$. VGG-15, on the other hand, exhibits more variation in validation UAR when distilled datasets are guided by different models, achieving its peak at $74.1\,\%$ UAR
under the guidance of its own architecture. This performance significantly surpasses the work presented in~\cite{10096757} ($p<0.01$ in a one-tailed z-test) and is comparable to ResNet-9’s performance trained on the full train dataset. Moreover, when trained on the distilled train plus validation dataset, representing a mere $14.9\,\%$ of the data samples, the test UAR experiences a decrease compared to the validation UAR. For instance, CNN-6, when trained with data distilled under ResNet-9’s guidance, yields a highest test UAR of $67.1\,\%$. Additionally, transferability of the distilled dataset has been observed. Specifically, CNN-6 and ResNet-9 perform similarly regardless of the guiding model for the distilled dataset, whereas VGG-15, when trained on data distilled by its own architectural style, outperforms the other models. This could be attributed to the inherent complexity and representational capacity of VGG-15, which may benefit more distinctly from data distilled within its architectural paradigm.

\vspace{-5pt}
\begin{table}[ht]
\caption{Model performances [Accuracy / UAR \%] on the DEMoS dataset. The models in the first column are used to guide the data distillation, while the models in the first row indicate those developed on the corresponding distilled data. IPC=$150$ means merely $28.2\,\%$ of the training data samples and $14.9\,\%$ of training plus validation data samples.}
\vspace{-10pt}
\begin{adjustbox}{width=1\columnwidth}
\begin{tabular}
{l|P{1.4cm}|P{1.4cm}|P{1.4cm}|P{1.4cm}|P{1.4cm}|P{1.4cm}}
\toprule
&\multicolumn{2}{c|}{\textbf{CNN-6}} & \multicolumn{2}{c|}{\textbf{ResNet-9}} & \multicolumn{2}{c}{\textbf{VGG-15}}\\
\cline{2-7}

IPC=150 &\emph{Val} & \emph{Test} & \emph{Val} & \emph{Test} & \emph{Val} & \emph{Test}\\
\hline
\textbf{CNN-6} &  74.8 / \textbf{74.9}  & 65.1 / 65.1 & 39.7 / \textbf{40.7}  & 39.9 / 39.9 & 51.8 / 50.2  & 50.8 / 51.3 \\
\hline
\textbf{ResNet-9} & 72.8 / 73.2 & 67.2 / \textbf{67.1} & 39.4 / 39.1  & 41.3 / \textbf{40.5} & 32.9 / 34.3 &  28.4 / 29.2 \\
\hline
\textbf{VGG-15} & 72.0 / 72.3 & 65.6 / 65.9 & 38.9 / 38.1  & 38.3 / 37.4 & 74.2 / \textbf{74.1} & 61.4 / \textbf{61.2} \\
\bottomrule

\end{tabular}
\label{tab:res}
\end{adjustbox}
\end{table}

\vspace{-20pt}

\section{Conclusions and Future Work}
\label{sec:conclusion}
In this work, we proposed a data distillation framework for Speech Emotion Recognition that leverages knowledge transfer from pre-trained teacher trajectories to generate a compact, synthetic dataset. Our experimental results demonstrated that the framework could achieve comparable performance to models trained on the full dataset, with the highest test unweighted average recall reaching $67.1\,\%$ using only $14.9\,\%$ of the data samples. Additionally, we observed notable transferability in the distilled dataset. In future work, we will investigate the application of more complex models (\eg wav2vec 2.0) to guide the data distillation process. We also intend to conduct an in-depth analysis of the characteristics of the generated distilled data to understand its underpinnings and potential for SER.

\bibliographystyle{IEEEtran}
\bibliography{mybib}

\end{document}